\documentstyle[11pt,aaspp4]{article}

\newcommand{\eq}{\begin{equation}}
\newcommand{\ee}{\end{equation}}

\newbox\grsign \setbox\grsign=\hbox{$>$} \newdimen\grdimen \grdimen=\ht\grsign
\newbox\simlessbox \newbox\simgreatbox
\setbox\simgreatbox=\hbox{\raise.5ex\hbox{$>$}\llap
     {\lower.5ex\hbox{$\sim$}}}\ht1=\grdimen\dp1=0pt
\setbox\simlessbox=\hbox{\raise.5ex\hbox{$<$}\llap
     {\lower.5ex\hbox{$\sim$}}}\ht2=\grdimen\dp2=0pt
\newcommand{\simgt}{\mathrel{\copy\simgreatbox}}
\newcommand{\simlt}{\mathrel{\copy\simlessbox}}

\lefthead{Hellsten \& Lin}
\righthead{Population II Star Formation}

\begin{document}

\title {On the Minimum Metallicity and Mass of the Population II Stars}

\author {Uffe Hellsten \& D. N. C. Lin\altaffilmark{1}}

\bigskip
\bigskip

\altaffiltext{2}{Astronomy Department, University of California, Santa Cruz,
CA, 95064\\Electronic mail: uffe@lick.ucolick.org, lin@lick.ucolick.org}

\begin{abstract}
Within collapsing protogalaxies, thermal instability leads to the formation of
a population of cool fragments which are confined by the pressure of a residual
hot background medium. The critical mass required for the cold clouds to 
become gravitationally unstable and to form stars is determined by both 
their internal temperature and external pressure.  We perform a systematic
study of the cooling properties of low-metallicity clouds, and we determine, 
for appropriate ranges of external pressure and metallicity, the minimum 
temperature clouds can attain prior to the formation of nearby massive stars.
The intense UV radiation of massive stars would photoionize the clouds 
and quench star formation.  We also determine the minimum metallicity 
these clouds must attain in order to form presently observable Population II
stars.  Based on our conclusion that low-mass stars cannot be formed 
efficiently in a metal deficient environment, we argue that brown dwarfs are
unlikely to be the main contributors to mass in the Galactic halo.

\end{abstract}

\section{Introduction}

During the collapse of a protogalactic cloud (PGC), density inhomogeneities 
and velocity variations lead to shocks which heat the gas to a temperature  
$T_{hot} \sim T_{vir}$, where $T_{vir}$ is the virial temperature of the 
galactic halo (Binney 1977; Rees \& Ostriker 1977; White \& Rees 1978).  In 
order for a PGC to collapse, its cooling time scale, $\tau_c$, must be shorter 
than the dynamical time scale, $\tau_d$, on which it can contract.  For PGC's
with masses comparable to the Galaxy, this condition is satisfied when their 
characteristic length scale $D<100$~kpc (Blumenthal {\it et al.} 1984).

Subsequent fragmentation of the PGC requires the growth of density 
inhomogeneities on a time scale, $\tau_g < \tau_d$. If the PGC 
is cold, gravitational instability causes perturbations with 
initial amplitude $\delta_0$ to become nonlinear when the system 
collapses by a factor $\sim \delta_0^{2/3}$ (Hunter 1962).  
This fragmentation process must proceed on scales down to individual
stars as a PGC contracts from an initial size $D=100$~kpc to the 
present size (a few kpc) of typical galaxies.  For such a small 
($\sim 10$) collapse factor, the amplitude of the perturbations 
must be a few percent for gravitational instability to trigger 
fragmentation of the PGC.

Large amplitude perturbations would be expected if bona fide normal
galaxies are formed as merger products of seed dwarf galaxies
with a size $\sim$ a few kpc and a mass $\sim 10^{6-7} M_\odot$
(Blumenthal  et al. 1984).  Residual gas within these building
blocks is expected to virialize prior to the merger events such that
$\delta_0$ on the protostellar cloud scale ($\sim 1$ pc) is likely 
to be small. Gravitational instability alone would not be adequate 
to induce the fragmentation of PGCs into protostellar objects.
However, in the limit that $\tau_c<\tau_d$, thermal instability can
lead to the rapid growth of perturbations from infinitesimal
$\delta_0$ to nonlinear amplitudes (Field 1965).  

For $T_{hot}\simgt10^5$ K, the dominant cooling mechanisms are 
bremsstrahlung, HeII line cooling, and H recombination cooling, and
$\tau_c$ increases with temperature.  In this case, any small temperature 
difference between the cooler perturbed regions and the background is 
amplified. Across the boundary of the perturbation, differential cooling 
leads to a pressure gradient which induces gas flow from the hot background 
towards the cooler perturbed regions, enhancing the local densities and 
leading to a runaway cooling process. When the cooling gas reaches
a temperature of $T \sim 16000 \, {\rm K}$ (see below), pressure equilibrium 
is re-established on all scales, resulting in a two-phase medium with 
density contrast inversely proportional to the temperature ratio. 
The residual halo gas (RHG) in the background remains at $T \sim T_{hot}$ 
with a density such that its thermal energy is lost on a time 
scale $\sim \tau_d$. Energy is lost from the RHG through both radiative 
cooling and thermal conduction between the RHG and the cold clouds 
(McKee \& Cowie 1977).  At a distance of $\sim 10$~kpc from the center of 
a Milky Way sized PGC, the energy balance implies $nT \sim 10^{3-4}$ (Lin 
\& Murray 1997).  

In the cooler clouds ($T = 10000 {\rm K} - 50000 {\rm K}$), the dominant 
cooling mechanism is line emission cooling from a trace amount of HI 
collisionally excited by free electrons. This process causes a virtually 
instantaneous cooling to $T \simlt 16000 {\rm K}$ (with $\tau _c \sim 10^4 
\, {\rm yrs}$ for $nT \sim 10^4$), below which the gas starts to recombine 
and a non-equilibrium treatment of the residual amount of free electrons 
becomes necessary. The equilibrium cooling function cuts off sharply at 
lower temperatures because the amount of free electrons is depleted rapidly,
causing the cooling to pause at $T \sim 10^4\, {\rm K}$. In reality, however, 
the cooling time scale is shorter than the H recombination time scale, 
resulting in a residual fraction of free electrons that can excite HI (and 
metals, if presents) and cause further line cooling. These electrons are also 
crucial for the formation of molecular hydrogen through the ${\rm H}^-$ - 
channel (e.g. Murray \& Lin 1992). In metal poor clouds, ${\rm H}_2$ cooling 
becomes dominant at $T \approx 6000 {\rm K}$ reducing the temperature to 
$\sim 10^2 {\rm K}$. When metals are present, OI, CI, CO, and grains provide 
additional cooling. In \S2, we perform a detailed study of the cooling of 
pressure confined clouds as well as clouds with constant densities.

The RHG also exerts a drag on the motion of the clouds as they 
are accelerated by the gravity of the Galactic halo.  The terminal speed of 
clouds with size L is $V_t \sim (f_n L/D)^{1/2} V_k$ where $f_n$ is the 
density ratio of the clouds to the RHG, and $V_k$ and $D$ are the velocity 
dispersion and size of the halo, respectively.  The motion of the clouds 
through the RHG also leads to mass loss due to the Kelvin-Helmholtz instability
(Murray et al. 1993), whose growth time scale ($\tau_{KH}$) is a few 
times $L/V_t$.  Because $\tau_{KH}$ increases with $\lambda$, the KH 
instability leads to fragmentation. The break down of the clouds increases 
their collective area filling factor and collision frequency.  A balance 
between disruption and coagulation establishes an equilibrium size 
distribution (Yorke, Lin \& Murray, 1997).

A lower limit to the size distribution of the clouds is set by their 
evaporation by the hot RHG.  In the high mass limit, the self-gravity of 
the clouds increases the central density and suppresses the Kelvin-Helmholtz 
instability. But at a critical mass $M_c \sim T^2 / (nT)^{1/2} M_\odot$, 
thermal pressure can no longer support the weight of the envelope (Bonnor 
1956), and the clouds undergo inside-out 
collapse (Shu 1977). During the collapse, although the Jeans mass decreases 
with density, it is larger than the mass contained inside any radius.  The
collapse is stable and does not lead to fragmentation without any further 
unstable cooling.  Thus, contrary to the opacity-limited fragmentation
scenario (Hoyle 1953; Low \& Lynden-Bell 1976), $M_c$ represents the
minimum mass for isothermal collapsing clouds (Tsai, in preparation).

In a metal-free environment, $T\sim 10^2$~K and $M_c \sim 10^2-10^3 M_\odot$. 
Thus, stars formed in a metal-poor environment are massive and short-lived, 
consistent with their rarity today.  Although lower $T$ and $M_c$ may be
attained in metal enriched clouds, low-mass star formation is regulated
by the emergence of massive stars which are copious sources of 
UV radiation.  Photoionization raises $T$ to above $10^4$~K and $M_c \sim 10^7 
M_\odot$. Small (a few $M_\odot$) heated clouds are stable and star 
formation is quenched.  Formation of low mass stars is possible provided
the clouds can cool to sufficiently low temperature prior to the onset of
nearby UV source.  In \S 3, we determine the timescale ($\tau_{10}$) for 
clouds to cool from $10^2$ to 10 K and compare it with the
dynamical timescale for relatively massive ($>20 M_\odot$) cool ($10^2$
K) clouds to collapse and form upper main sequence stars.  We also  
evaluate the minimum value of $M_c$ as a function of $nT$ and [Fe/H].
Finally in \S4, we discuss the implication of our results for the formation of
Population II stars in globular clusters.

\section{Cooling of low-metallicity clouds}
We are interested in following a gaseous region as it cools and contracts
due to the pressure of a surrounding hot, confining gas. For temperatures 
in the range $\log{T} \leq 4.2$ helium is for all practical purposes fully 
recombined and does not contribute to the density of free electrons.
To calculate the time-dependent, non-equilibrium fractions of electrons 
and ${\rm H}_2$ in the cooling gas we integrate the rate equations for
the following network of reactions:

\begin{eqnarray}
{\rm H}^+  +  e^- & \rightarrow & {\rm H}  +  \gamma \nonumber \\
{\rm H}  +  e^- & \rightarrow & {\rm H}^+  +  2e^- \nonumber \\
{\rm H}  +  e^-  & \rightarrow & {\rm H}^-  +  \gamma \nonumber \\
{\rm H}  +  {\rm H}^- & \rightarrow & {\rm H}_2  +  e^-  \\
{\rm H}^-  +  e^- & \rightarrow & {\rm H}  +  2e^- \nonumber \\ 
{\rm H}^-  +  {\rm H}^+ & \rightarrow & 2 {\rm H} \nonumber \\
{\rm H}_2  +  {\rm H}^+ & \rightarrow & {\rm H}^+_2  +  {\rm H} \nonumber \\
{\rm H}_2  + e^- & \rightarrow & 2 {\rm H}  +  e^- \nonumber
\end{eqnarray}

These are the reactions included in the "minimal model" suggested by 
Abel et. al. (1996), from which we have taken the expressions for the
reaction rates as a function of temperature. This model agrees to within a few 
percent with a more elaborate model featuring 19 reactions. 
For the cooling resulting from excitation of the rotational and vibrational
states of ${\rm H}_2$ by ${\rm H}-{\rm H}_2$ collisions we adopt the expression
given in Tegmark et. al. (1996).

We have included the solutions of the rate equations from the above network
and the resulting ${\rm H}_2$ cooling into a driver code for the 
photoionization code CLOUDY 90 (Ferland 1996). CLOUDY is used to calculate 
the HI line cooling as well as cooling from line excitation of metals, as 
a function of density, temperature, and the non-equilibrium electron density. 
To follow the time-dependent cooling of the gas we also need to know the 
evolution of the gas density.  Throughout this paper we will describe the 
gas density by the parameter $n$, the total number density (in ${\rm cm}^{-3}$)
of hydrogen atoms in a gas with hydrogen and helium mass fractions X=0.76 
and Y=0.24, respectively. The units of the quantity $nT$, which can be viewed 
as a measure of the pressure of a hot, confining medium, is ${\rm K}\, 
{\rm cm}^{-3}$. Two limiting cases will be considered here: 1) $nT= 
{\rm constant}$ (isobaric) and 2) $n = {\rm constant}$ (isochoric). The 
first case mimics the compression of a cooling gas cloud by a surrounding, 
confining medium and applies to regions cooling coherently on length scales 
small enough for the sound crossing time to be shorter than the cooling time 
scale. For larger regions (of the same density), the gas cools faster than 
it can adjust to hydrostatic equilibrium, and is better described by the
isochoric cooling model. 

The calculations are started at $\log{T} = 4.2$ with the equilibrium value
of the hydrogen ionization fraction $x=0.54$ and a negligible fraction of
molecular hydrogen. The initial hydrogen number density $n_0$ and the
metallicity [Fe/H] of the gas are treated as free parameters, and we
will examine the properties of the cooling models for $(nT)_0$ ranging
from $10^3$ to $10^5$ and [Fe/H] from -2 to $-\infty$. Only C and O
turn out to be important coolants, and we assume [C/Fe]=0 and [O/Fe]=0.5, 
corresponding to the observed enrichment pattern in low-metallicity
regions (eg. Wheeler, Sneden \& Truran 1989). Dust grains are assumed
to be absent. They are likely to play a negligible role in such 
low-metallicity systems, and it is not possible to model the dust
formation process within these cooling clouds in any reliable way.  

For illustrative purposes we plot in Figure 1 the value of $dT/dt$ as a 
function of $T$ for [Fe/H]= -3.5, -3, and  -2.5 for the isobaric and isochoric
models with $(nT)_0 = 10^4$. The contribution from ${\rm H}_2$ is plotted 
separately. It is seen that for ${\rm [Fe/H]} \simlt -2$, ${\rm H}_2$ is the 
dominant coolant in the entire temperature interval $\log{T} \approx 2.4-3.8$. 
The most important {\em metal} contributor in this interval is OI. The 
${\rm H}_2$ cooling rate cuts off steeply for $T \simlt 10^2 {\rm K}$, which is
the lowest temperature to which a primordial gas can radiatively cool. For 
$10\, {\rm K} < T < 100\, {\rm K}$, excitation of the CI (369 $\mu m$, 609 $\mu m$)
transition (mainly by H-CI collisions) is the dominant cooling mechanism, 
except for the bump in the cooling function at around $10^2 \,{\rm K}$, 
which is caused by CO.  Also note that in the metal deficient limit, cooling 
below $T \sim 10^2$ K is thermally stable since $T / (dT/dt)$ decreases with 
$T$.  But for [Fe/H] = -2, $d T / dt \propto T$ and 
the cooling is marginally thermally unstable.  

In Figure 2, the evolution of $T$ for [Fe/H] = -4, -3, and -2  is shown for 
the isobaric model with $nT=10^4$. This is an adequate description for the 
cooling of clouds of sizes
\begin{equation}
S < S_c \equiv c_s T/ (dT/dt),
\label{scrit}
\end{equation}
where $c_s=\sqrt{kT/\mu m_H}$ is the sound speed in the cloud. As expected, 
the cooling of a metal free cloud is stalled at $\sim 10^2 \,{\rm K}$ (for 
the time interval considered, the [Fe/H] = -4 model is practically identical 
to a zero-metallicity model). Although it is possible to cool the gas to 
10 K for [Fe/H] = -3, it takes more than $3 \times 10^7$ yrs.
As we will argue in the following section, long-living stars with masses
${\rm M} \simlt 1 {\rm M}_{\odot}$ will only be able to form in an environment
where $T \sim 10\, {\rm K}$, whereas short-lived, massive stars can also form
in a $10^2 \,{\rm K}$ environment. 

\section{Minimum stellar mass}

In \S1, we indicated that the mass spectrum of the cool clouds is determined
by an equilibrium between fragmentation due to their interaction with RHC
and cohesive collisions.  These processes lead some clouds to grow.  When 
an isothermal cloud approaches a critical mass $M_c$, it becomes 
self-gravitating and undergoes collapse. The critical value is 
\begin{equation}
M_c \approx 27 M_\odot \, f\frac{T^2}{\mu ^{3/2}{(nT)}^{1/2}},
\label{mcrit}
\end{equation}
where the factor $f$ is equal to unity in the classical Bonnor-Ebert analysis 
of the stability of isothermal gas spheres. If, however, the cloud is 
dynamically perturbed with a finite amplitude, collapse is triggered in 
less massive clouds, and $f<1$.  Nonlinear perturbations may be induced by
collisions between marginally stable clouds.  The mean molecular weight, 
$\mu$, is around 1.2 in a neutral medium of H and He, and 2.3 if most of 
the hydrogen is molecular. In practice, we will absorb the uncertainty
in $\mu$ into the factor $f$. As will be demonstrated, our inferred 
minimum metallicity for pop II stars is quite insensitive to variations in 
that factor. It should be noted that as a cloud builds up and approaches the 
Bonnor-Ebert mass, it also acquires a density gradient. In equation (\ref
{mcrit}), $n$ denotes the mean gas density within the Bonnor-Ebert radius. 

Let us now look, for a moment, at the properties of present star-forming 
regions within our Galaxy. Stars of a wide range of masses are known to be 
forming within dense molecular cores embedded in giant molecular clouds. 
In these regions, the physical conditions are $T \sim (10-20) \, {\rm K}$ 
and the thermal pressure $nT \sim 3 \times 10^5$ (Lada 1993). From equation 
(\ref{mcrit}) we find that the collapse of long-lived, low mass stars with 
$M \simlt 1 M_{\odot}$ can be triggered in such regions, provided that $f 
\simlt 0.3$. Since the magnetic pressure is comparable to $nT$ in these 
cores (Myers 1993), similar critical mass for gravitational unstable magnetized 
clouds is inferred (McKee et al. 1993).  For larger temperatures, it 
becomes increasingly more difficult (i.e. requiring lower $f$ and hence larger
dynamical perturbations) to form stars of a given mass. We thus expect most,
if not all, of the low mass stars to be formed within regions of the lowest 
attainable temperatures, $T=(10-15) \, {\rm K}$, while stars of increasingly 
higher mass can collapse at increasingly higher temperatures. 
In the model calculations to be presented below, we will assume that 
$M_c(T=10 {\rm K}) < 1 M_{\odot}$ and we will adjust the ratio $f \mu ^{-3/2}$ 
accordingly, for a range of $nT$.

The characteristic dynamical timescale for a cloud is given by 
\begin{equation}
\tau_d \equiv (G \rho) ^{-1/2} = 8.3 \times 10^7 \, {\rm yrs} (T/nT)^{1/2}, 
\label{tgrav}
\end{equation}
which is the timescale for a cloud to collapse after it becomes marginally 
unstable. For the clouds that we are considering, it is of order a few million
years.  In reality, both radiation feedback and rotation can slow down the 
collapse (Stahler, Shu, \& Taam 1980, Yorke et al. 1995). On the other 
hand, the high density {\em core} of the unstable cloud collapses on a shorter
timescale than that given by equation (\ref{tgrav}).

Nevertheless, we adopt the assumption that upper main sequence stars 
(with $M>M_{\rm UV}$) begin to release intense UV radiation after a time 
$\tau_e = \tau_d$ has elapsed from the epoch $t(M_{\rm UV})$ at which
$M_c$ in the clouds is reduced below $M_{\rm UV}$.  From (\ref{mcrit}) 
and (\ref{tgrav}), we find
\begin{equation}
\tau_e = 3.6 \times 10^7 \, {\rm yrs} \; f^{-1/4}\, {M_{\rm UV}}^{1/4}\, \mu ^{3/8}\, (nT)^{-3/8}.
\end{equation}
We will use the value $M_{\rm UV}=20 M_{\odot}$ in all the calculations below
because $\tau_e$ depends only weakly on this characteristic mass of strong UV
emitters.

A minimum temperature 
\begin{equation}
T_{\rm min} \equiv T(t(M_c=M_{\rm UV})+\tau_e). 
\end{equation}
is attainable in a region before it is re-heated by the UV photons
from emerging massive stars.  The heating and photoionization of the clouds 
stabilize them against collapse. The value of $T_{\rm min}$ is a function 
of [Fe/H] and $nT$.  The corresponding minimum mass of stars that can form is, 
\begin{equation}
M_{\rm min}=M_c(T_{\rm min}). 
\end{equation}
Figure 3 shows $M_{\rm min}$ as a function of metallicity for $\log{nT}=3,4$, 
and 5. For each $nT$, $f$ has been chosen so that $M_c(10 {\rm K})/M_{\odot}=
0.8, 0.4$, and 0.2, requiring values of $f$ in the interval [0.001;0.2] 
(for $\mu =2.3$). Specifically, f=(0.016,0.008,0.004) for $\log{nT}=3$, f=(0.05,0.025,0.013) 
for $\log{nT}=4$, and f=(0.16,0.08,0.04) for $\log{nT}=5$ respectively.  It is evident 
from Figure 3 that there is a limit in metallicity below which the formation 
of long-lived stars (with $M < 0.8 \, M_{\odot}$) is quenched. For $\log{nT}
=3,4,5$, this limit is at [Fe/H] $\simeq$ -2, -2.5, and -3, insensitive to 
the exact choice of $f$. 
 
As an alternative way of presenting these results we plot $T_{\min}$ as a
function of metallicity in Figure 4. For clarity we only show the
models for which $M_c(10 \, {\rm K}) = 0.4 \, M_{\odot}$ here. For low 
metallicities, the cooling is too weak to cause any significant temperature
reduction below $T(M_{UV})$ during the time $\tau _e$, and the curve is 
flat. The transitions to $T_{\min}= 10 \, {\rm K}$ occur sharply around 
[Fe/H]= -2, -2.5, and -3, the lower limits in metallicity also inferred 
from Figure 3.

In the above calculations, the isobaric cooling model was used.  This 
assumption is valid provided $M_c$ is less than the mass contained within 
$S_c$ (equation (2)): 

\begin{equation}
M_c \, < \,  \frac{4}{3} \pi S_c^3 \, \rho \, \propto \, (nT) \, T^{7/2} / (dT/dt)^3 \, .
\label{isobarcond}
\end{equation}

Figure 5 shows the mass within $S_c$ and $M_c$ as a function of $T$ for 
various metallicities in the model with $\log{nT}=4, M_c(10 \, {\rm K}) 
= 0.8 M_{\odot}$.  It is clear that the condition (\ref{isobarcond}) is 
satisfied for all metallicities [Fe/H] $< -2.5$, so the calculations 
that determine the minimum metallicity of low-mass stars in Figures 3 
and 4 are self-consistent. Similar considerations apply to the other 
models presented in these figures. For higher metallicities, i.e. on 
the flat part of the $M_{\rm min}$-[Fe/H] curves, the isobaric 
approximation gets worse, and the minimum mass becomes uncertain.  
Also, dust may play an increasingly important role as the heavy 
element abundance increases. 

\section{Discussions}

Globular clusters contain the oldest and most metal deficient stars in 
the Galaxy. The results presented here are particularly relevant to the 
history of star formation in these stellar systems.  The presence of
low-mass cluster stars today clearly indicate that $M_c < < 0.8 M_\odot$.
The expression in (\ref{mcrit}) indicate that $M_c$ depends on $T$ much more 
sensitively than $nT$, although the values of both $T_m$ and $M_c(T_m)$ 
vary more rapidly with $nT$ (see Figure 3).  In \S1, we presented theoretical
arguments to estimate $nT \sim 10^4$.  Here we infer the value of $nT$ 
of proto globular cluster clouds (PCC's) from the current properties of 
globular clusters, averaged over their half-mass radius ($r_h$). If both
PCCs and the individual clouds are pressure confined, $nT$ is the same for 
all the entities in the PGC including the RHC.

During globular clusters' dynamical evolution, their density ($n$) and 
velocity dispersion at $r_h$ do not change significantly after the epoch 
of their formation.  However, the extrapolation of the physical condition 
of PCC's to the stage prior to star formation is highly uncertain.  
If, after their formation, the stars undergo collapse and virialization 
from rest, the clouds' initial radii ($r_i$) would be $\sim 2 r_h$.  Larger
ratios between $r_i$ and $r_h$ would be expected if star formation 
requires dissipative collisions and coagulation of substellar fragments 
(Murray \& Lin 1996). But $r_i$ is unlikely to be larger than the tidal 
radii of the PCC's, which are typically only a few times larger than 
the present values of $r_h$.  Thus, the initial density of 
PCC's may be 1-3 orders of magnitude smaller than the average cluster density 
at $r_h$ today.  Based on the present velocity dispersion of the clusters, 
we infer the initial temperature of the PCC's to be $\sim 10^4$~K, comparable
to that expected if they were photoionized.  From these estimates, we infer
$nT \sim 10^{2-5}$, and that $nT \propto D^{-3}$ where $D$ is the cluster's 
distance to the Galactic center.  In accordance with the pressure confinement 
scenario, both the magnitude and the spatial dependence of PCC's $nT$
are consistent with those expected for the RHG (Murray \& Lin 1992).

From these results, and the cluster metallicities, we can also estimate
the cooling time scale and dynamical time scale of the PCC's.  The ratio of 
these time scales increases from $\sim 10^{-4}$ near the Galactic bulge to 
$\sim 1$ at $\sim 100$ kpc.  In most PCC's, thermal equilibrium is only 
possible in the presence of external UV photons with a flux comparable to 
that required by self regulated star formation in the halo (Lin \& Murray 
1992). An independent upper limit on $nT$ is inferred for the PCC's from 
the requirement that their column density must be less than that would 
self shielded against the external UV flux (Hellsten, Lin, \& Murray 1997).  
If the value of $nT$ is too small, the PCC's would be confined by self 
gravity, in which case the heating and cooling equilibrium is no longer stable
(Murray \& Lin 1992). Marginal self gravity provides a favorable condition
for PCC's to retain most its gas content against the ram pressure stripping by
the RHG (Murray  et al. 1993) and for efficient star formation to be 
triggered by relatively small amplitude perturbations.  

Based on the considerations, we estimate $nT \sim 10^{2-5}$.  These values 
of $nT$ are used in Figs 1-4.  For this range of $nT$, the results in Figure 3 
indicate that the formation of the low-mass stars would be quenched by nearby 
massive stars unless the GCs were formed from gas that had already been 
pre-enriched to a metallicity [Fe/H]$\sim -3$ to -2. At least two general 
observational results on the metallicity of the globular clusters in the 
halo of our Galaxy can be easily reconciled with our scenario
on the formation of low mass stars. First, the inferred homogeneous 
metallicity among individual stars within a globular cluster would be 
expected if the GC is formed out of a well-mixed, pre-enriched PGC rather 
than being chemically self-enriched (Murray \& Lin 1993). Second, the 
metallicity distribution of globular clusters shows a cut off below [Fe/H] 
$\approx -2.5$ (e.g. Ryan \& Norris 1991), which is what would be 
expected if these low-metallicity clusters were formed in an environment 
with $\log{nT} \sim 4$.  Clusters with [Fe/H] $<$ -2.5 may have formed   
in this environment at these early epochs, but these cluster would contain 
mostly massive stars.  Mass loss associated with the rapid evolution 
of massive stars can lead to the disruption of these clusters.
In addition, $M_c$ represents the minimum stellar mass. For metal 
deficient clouds with [Fe/H] $<$ -2.5, $M_c > 1 M_\odot$ such that 
all cluster stars would have evolved off the main sequence phase within the
Hubble time.
  
Our results are also consistent with the rarity of extremely metal-deficient 
low-mass stars in the Galactic halo.  In the outer region of the halo where 
$nT \sim 10^3$, relatively large values of [Fe/H] ($>-2$) are needed for 
the formation of low $M_c$ stars.  The generation of such large amount of 
heavy elements within one free-fall timescale of the PGC would require 
the coexistence of $> 10^5$ upper main sequence (earlier than O5) stars.  
The filling factor of their Stromgren sphere in the PGC would greatly 
exceed unity such that their UV flux should have quenched star formation 
well below this level.  Based on these considerations, we do not expect the
formation of low-mass stars (including brown dwarfs) to be an efficient 
process in the outer regions of the Galactic halo.  Thus, we suggest the 
stellar or substellar objects are probably not the main contributors to the 
mass in the Galactic halo.

On the observational side, a search for stellar objects (MACHO) has led to the
identification of several microlensing events toward the direction of the 
Large Magellanic Clouds (LMC) (Alcock et al. 1996).  The characteristic mass of 
the lensing objects is $\sim 0.5 M_\odot$. If the angular distribution of these 
objects is uniform, the observed events would not only be sufficient to 
account for the mass inferred for the Galactic halo, but also contradict
our main conclusion that low mass stars cannot form efficiently in the Galactic
halo.  However, Zhao (1997) suggested these lensing objects in the direction
of the LMC may be old stars which were tidally torn form the LMC by the 
Galaxy.  A recent survey provided observational support for the existence 
of this group of stars between the Galaxy and the LMC (Zaritsky \& Lin 1997).  
The absence of similar stars in other directions suggests that 1) the LMC 
direction along which lensing objects are found may be special and 2) the  
density of stellar objects in the Galactic halo remain uncertain.

\acknowledgements
We thank Drs. P.H. Bodenheimer, A. Burkert, L. Hernquist, \&   
S.D. Murray for useful conversations.  This work is supported 
by NSF and NASA through grants AST-9315578, ASC 93-18185, NAGW-4967, NAG5-3056,
and NCC-2-649.
UH acknowledges support by a postdoctoral research grant from the 
Danish Natural Science Research Council.

\centerline{Figure Captions}

\figcaption{Cooling rates $dT/dt$ for isobaric (solid lines) and
isochoric (dashed lines) models. $nT=10^4 \, {\rm K}/{\rm cm}^3$ and
[Fe/H] = -2.5, -3, and -3.5} 

\figcaption{Temperature as a function of time for isobarically cooling gas with 
$nT=10^4 \, {\rm K}/{\rm cm}^3$ and [Fe/H]= -2, -3, and -4.}

\figcaption{Minimum mass of stars that can form within regions with $\log{nT}=3,4,5$
as a function of [Fe/H]. For each value of nT, the calculation is performed for the
three values of $M_c(10 \, {\rm K})/M_{\odot} =0.8,$ 0.4, and 0.2.} 

\figcaption{Minimum temperature that a star-forming region can acquire prior to
reheating by UV radiation from massive stars. Values of $nT$ are indicated on the
figure. For the models shown, $M_C(10 \, {\rm K}) = 0.4 \, M_{\odot}$. } 

\figcaption{Mass contained within the radius $S_c$ of equation (2) as a function of
temperature for various metallicities in a model with $\log{nT}=4$ and 
$M_c(10 \, {\rm K}) = 0.08 M_{\odot}$ (solid lines). Also shown is $M_c(T)$ 
(dashed line).}

\clearpage

\plotone{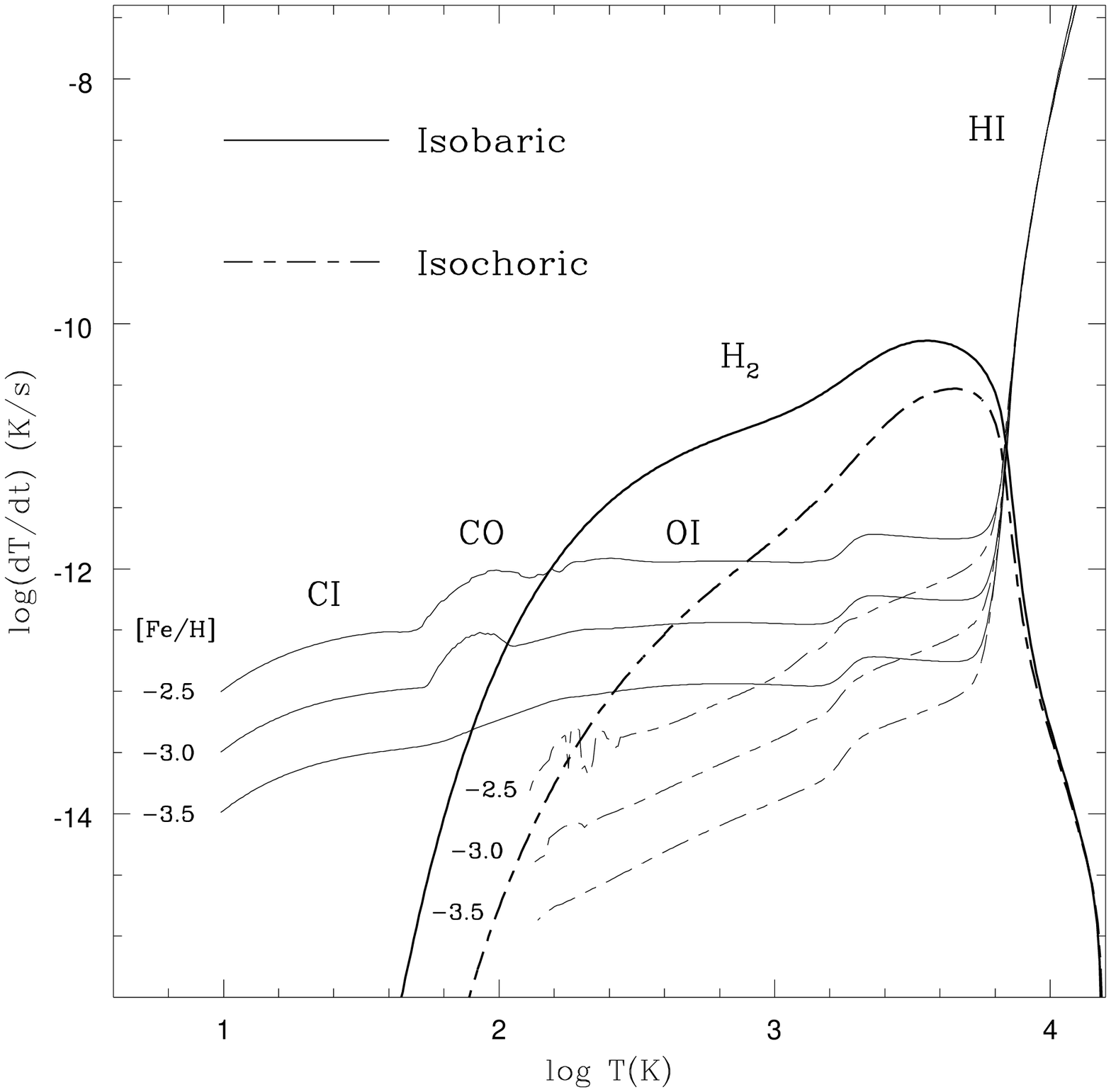}

\clearpage

\plotone{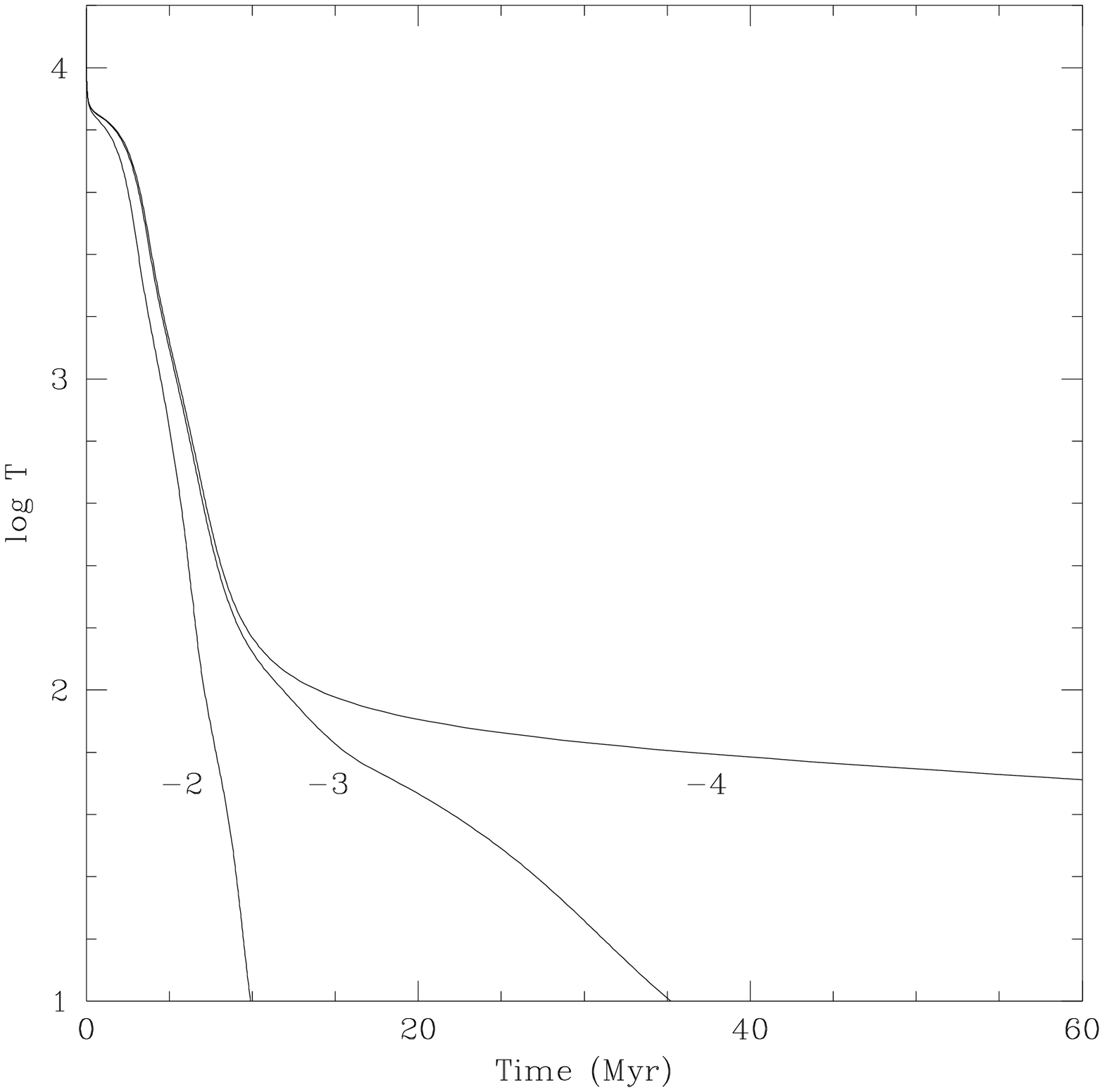}

\clearpage

\plotone{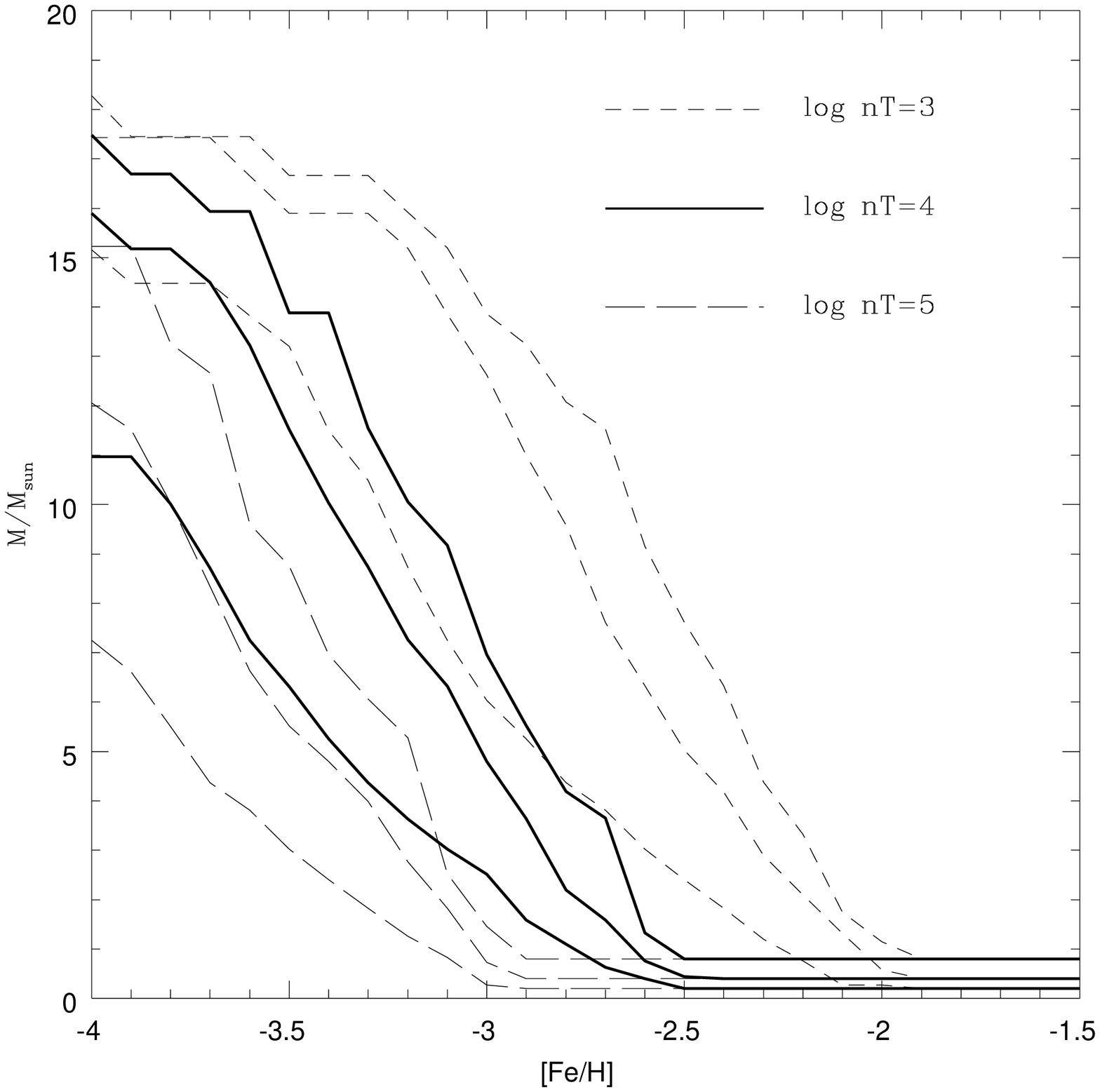}

\clearpage

\plotone{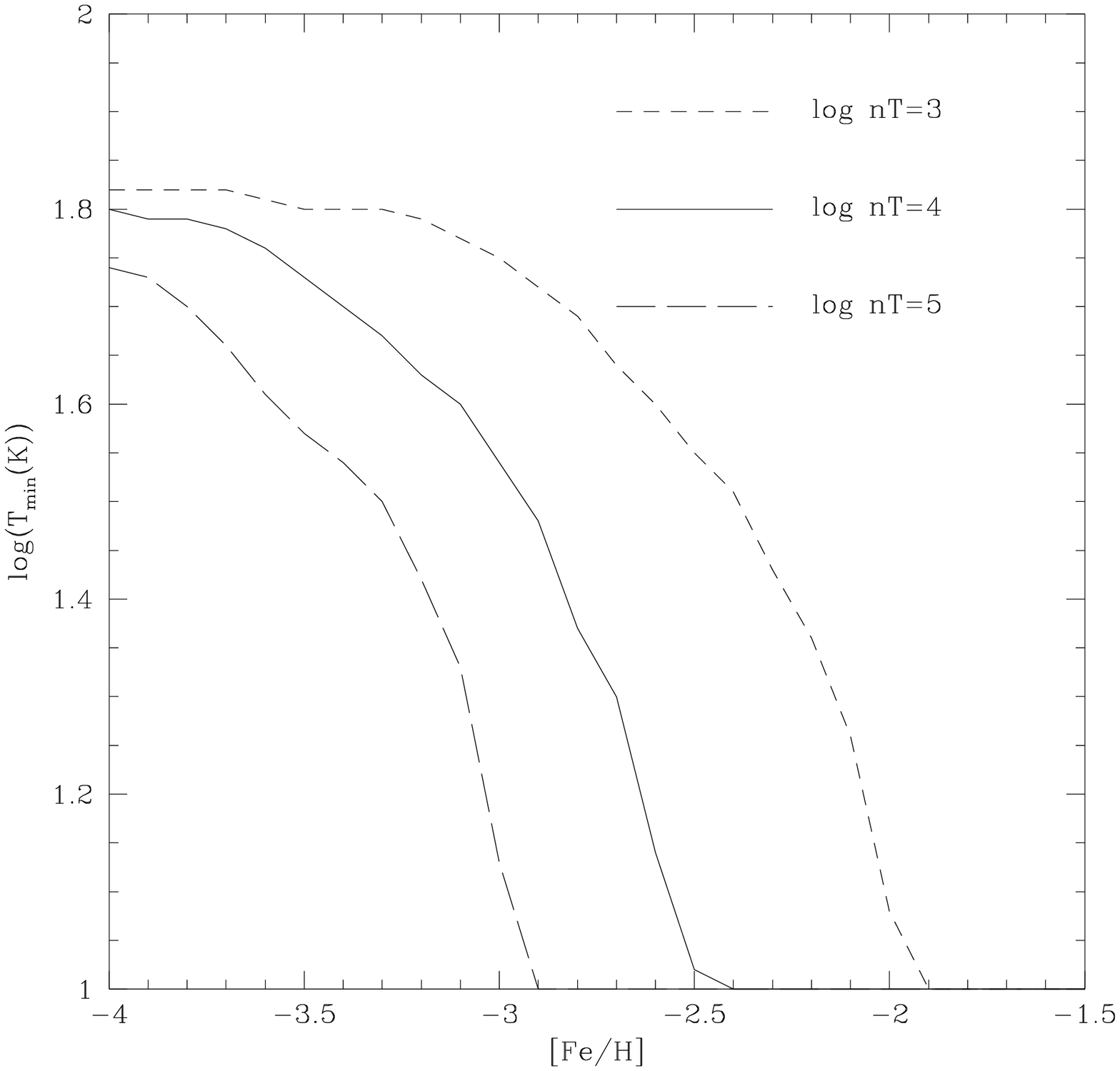}

\clearpage

\plotone{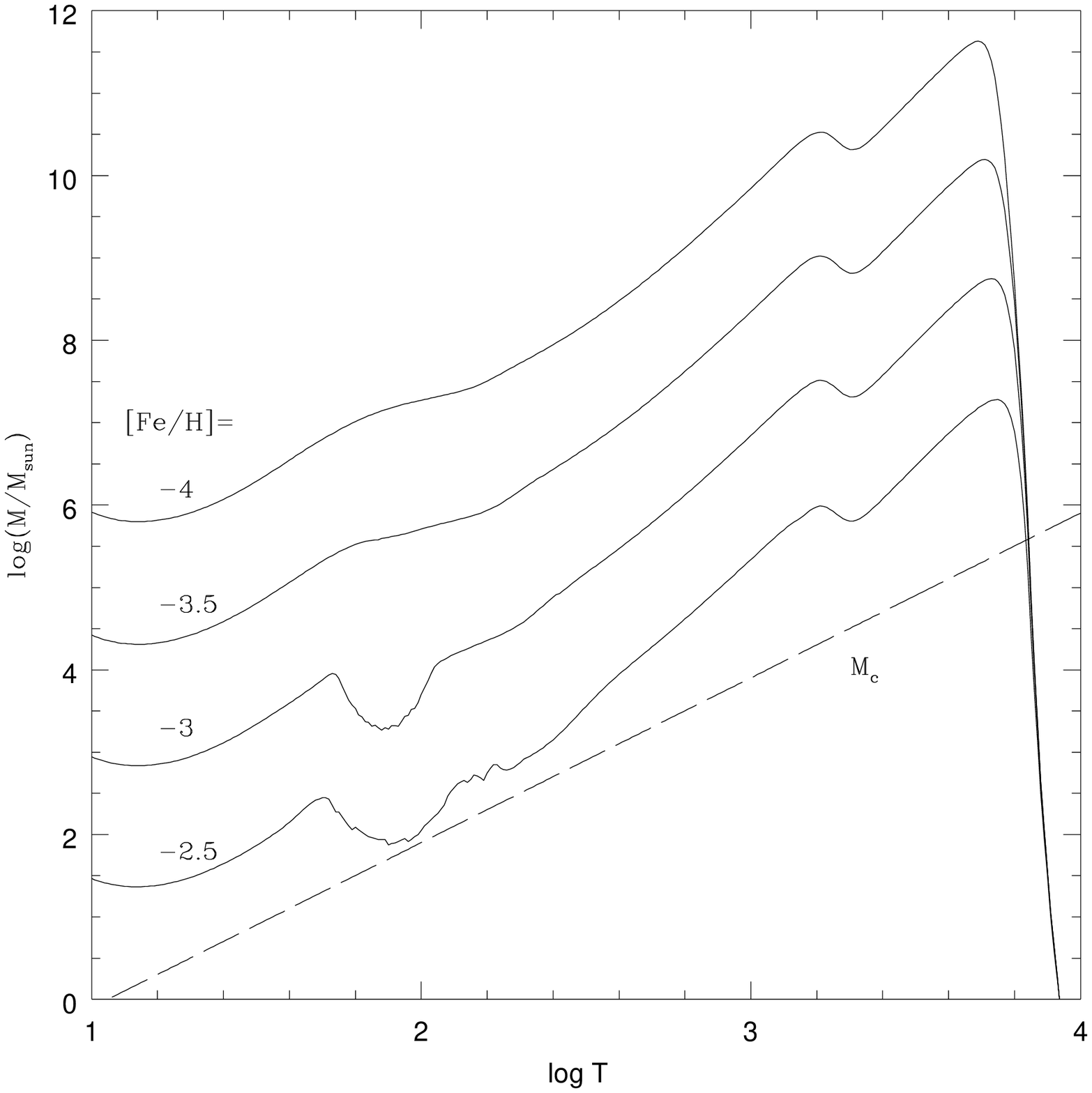}

\end{document}